\documentclass[iop,apj,numberedappendix]{emulateapj}
\usepackage[latin1]{inputenc}
\usepackage{amssymb}
\usepackage{amsmath}
\usepackage{graphicx}
\usepackage{xspace}
\usepackage{natbib}
\usepackage{url}
\usepackage{paralist}
 \usepackage{aasmacros}
\usepackage{multirow}

\slugcomment{ }

\shorttitle{The rhythm of Fairall~9 -- Spectral variability}
\shortauthors{Lohfink et al.}

\begin{document}

\title{The rhythm of Fairall~9 -- \\I: Observing the spectral variability with XMM-Newton and NuSTAR}


\author{A.M. Lohfink\altaffilmark{1}, C.S. Reynolds\altaffilmark{2}, C. Pinto\altaffilmark{1}, W. Alston\altaffilmark{1}, S.E. Boggs\altaffilmark{3}, F.E. Christensen\altaffilmark{4}, W.W. Craig\altaffilmark{3,5}, A.C. Fabian\altaffilmark{1}, C.J. Hailey\altaffilmark{4}, F.A. Harrison\altaffilmark{6}, E. Kara\altaffilmark{1}, G. Matt\altaffilmark{7}, M.L. Parker\altaffilmark{1}, D. Stern\altaffilmark{8}, D. Walton\altaffilmark{8}, W.W. Zhang\altaffilmark{9}
}
\email{alohfink@ast.cam.ac.uk}
\altaffiltext{1}{Institute of Astronomy, University of Cambridge, Madingley Road, Cambridge CB3 0HA, UK}
\altaffiltext{2}{Department of Astronomy, University of Maryland College Park, College Park, Maryland 20742, USA}
\altaffiltext{3}{Space Sciences Laboratory, University of California, Berkeley, 7 Gauss Way, Berkeley, CA 94720-7450, USA}
\altaffiltext{4}{Danish Technical University, DK-2800 Lyngby, Denmark}
\altaffiltext{5}{Lawrence Livermore National Laboratory, Livermore, CA, USA}
\altaffiltext{6}{Cahill Center for Astronomy and Astrophysics, Caltech, Pasadena, CA 91125, USA}
\altaffiltext{7}{Dipartimento di Matematica e Fisica, Universit\`{a} degli Studi Roma Tre, via della Vasca Navale 84, I-00146 Roma, Italy}
\altaffiltext{8}{Jet Propulsion Laboratory, California Institute of Technology, 4800 Oak Grove Drive, Pasadena, CA 91109, USA}
\altaffiltext{9}{X-ray Astrophysics Laboratory, NASA/Goddard Space Flight Center, Greenbelt, MD, 20771, USA}


\begin{abstract}
We present a multi-epoch X-ray spectral analysis of the Seyfert~1 galaxy Fairall~9. Our analysis shows that Fairall~9 displays unique spectral variability in that its ratio residuals to a simple absorbed power law in the 0.5-10\,keV band remain constant with time in spite of large variations in flux. This behavior implies an unchanging source geometry and the same emission processes continuously at work at the timescale probed. With the constraints from \textit{NuSTAR} on the broad-band spectral shape, it is clear that the soft excess in this source is a superposition of two different processes, one being blurred ionized reflection in the innermost parts of the accretion disk, and the other a continuum component such as spatially distinct Comptonizing region. Alternatively, a more complex primary Comptonization component together with blurred ionized reflection could be responsible.    
\end{abstract}



\section{Introduction}\label{intro}

The radiation from the innermost regions of active galactic nuclei (AGN) is primarily emitted in the X-ray band, making X-ray spectroscopy a prime tool for studying this region. Moreover, the variability of the radiation can provide additional diagnostics. Flux and spectral variations in AGN are usually nothing out of the ordinary. The drivers and nature of this variability are known to depend on the timescale of the variability. For example, changes in the accretion rate govern the longest timescales, and lead the continuum to become softer when brighter in the X-ray band \citep{Markowitz2003}. Absorption changes can drive strong variability, with timescales and spectral signatures that depend on the type of absorber. In general, all timescales depend on the black hole mass \citep[see][and references therein]{McHardy2006,Ponti2012}, where the timescales are longer with increasing black hole mass.  

In this light, the X-ray spectral variability of Fairall~9 stands out. This luminous Seyfert 1 galaxy ($z=0.047$), which has been well-studied throughout the years, is renowned for its lack of a warm absorber \citep{Reynolds1997,Emmanoulopoulos2011} and its simple X-ray spectrum \citep{Walton2012}. The spectrum is well described by a powerlaw continuum with cold and ionized reflection \citep{Emmanoulopoulos2011,Walton2012}, possibly with an additional soft X-ray continuum component \citep{Lohfink2012a,Patrick2011}. In particular, from a multi-epoch spectral analysis including \textit{XMM} data from 2009 and \textit{Suzaku} data from 2007 and 2010, \citet{Lohfink2012a} conclude that the spectral modeling of the soft excess is ambiguous. Both a multi-zone ionized reflection model and a model with an additional Comptonization component with a single zone ionized reflector describe the data. The accretion disk parameters are dependent on the model chosen. For example the black hole spin is determined to be $0.97_{-0.01}^{+0.02}$ for a reflection-only soft excess and $0.52_{-0.15}^{+0.19}$ for a Comptonization soft excess. The high black hole mass of $(2.55\pm0.56)\times10^8\,M_\odot$ \citep{Peterson2004} suggests low amplitude short-term variability, and indeed the source does not display strong intra-observation variability on hour to day timescales \citep{Emmanoulopoulos2011}. The discovery of surprisingly rapid X-ray flux dips in 10\,years of \textit{RXTE} monitoring however has put this picture of low amplitude variations  on short timescales into question \citep{Lohfink2012}. During a dip the 2-10\,keV flux decreases by a factor of two or more. These flux dips happen on a timescale of a few days, so slightly longer than the length of a typical long X-ray observation.  While it was not possible to determine the reason for the flux dips from the \textit{RXTE} monitoring, several possibilities have been suggested: inner disk or coronal instabilities, absorption from Compton-thick clumps in the torus, or even a failed jet cycle. 

The aim of this paper is to study the broad-band spectral shape and the nature of the soft X-ray excess in Fairall~9. To do so we have obtained new high quality \textit{XMM} CCD observations combined with \textit{NuSTAR} hard X-ray coverage targeted at answering these questions. In this paper we present a spectral analysis of these data. A future paper will discuss the variability and timing properties. The paper is organized in the following way; in \S\ref{reduction} we discuss the data reduction followed by a description of the results from the spectral analysis in \S\ref{results}. Before finally discussing the results (\S\ref{discuss}), we briefly show in \S\ref{previous} how our new insights are consistent with the spectra already presented in \citet{Lohfink2012a}. 

\section{The observing campaign}

\begin{figure}[ht]
\includegraphics[width=\columnwidth]{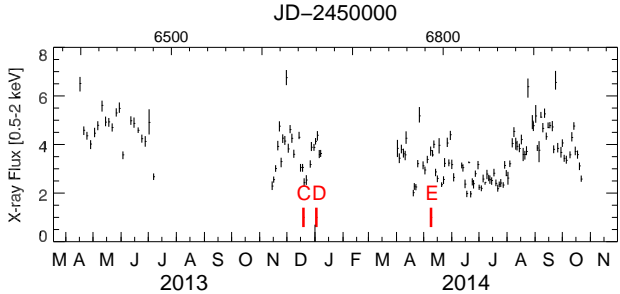}
\vspace{-8\baselineskip}
\caption{Overview \textit{Swift} X-ray light curve of the 0.5-2\,keV X-ray flux in units of $10^{-11}$\,erg/s/cm$^2$  indicating the flux state of the three newest \textit{XMM} observations (XMMC, XMMD, XMME) as red bars. It is apparent that XMMC was taken at very low flux, while XMMD and XMME were at high flux.}\label{light_f9}
\end{figure}

To learn more about the X-ray flux dips, we designed an observing campaign to provide a detailed spectral and
 temporal study of the source. It consists of three parts, starting with a \textit{Swift} \citep{Gehrels2004}
 monitoring campaign to observe Fairall~9 in the optical, UV, and X-ray bands for several months in 2013 and 2014 with an average cadence of 2-days
  to search for dips. The earliest part of this monitoring which ranged from April to June 2013 was presented in
  \citet{Lohfink2014}. In that study we focused on the connection of the X-ray variability to the UV variability and found correlated optical/UV variability on all time scales sampled. We reduced the newly added data in a similar fashion to that from the pilot study. The resulting X-ray light curve is shown in Fig.~\ref{light_f9}. The dips are immediately apparent in a visual analysis of the X-ray light curve (Fig.~\ref{light_f9}). Based on this monitoring, we triggered two \textit{XMM} observations, one in a dip and one out of a dip constituting the second part of our campaign. These observations are indicated as vertical red bars in the light curve. The third part of our campaign (indicated by the third red bar in the light curve) corresponds to a joint \textit{XMM}-\textit{NuSTAR} observation meant to reveal the exact broad-band spectral shape of Fairall~9. This observation was also taken at high flux, although not as high as the non-dip observation. The flux variability is also apparent from the unfolded \textit{XMM} spectra shown in Fig.~\ref{spectra_unfolded}\footnote{The spectra were obtained by assuming a diagonal response matrix.}. With three observations taken at different flux states, we have a dataset that is well-suited to study possible flux-related spectral changes. 
 
\section{High Quality Spectroscopic Data and Data Reduction}\label{reduction}
\subsection{Available High-Quality Spectral Datasets}

Fairall~9 is a well-studied AGN; an overview of the existing data from \textit{XMM}, \textit{Suzaku} and \textit{NuSTAR} is given in Table~\ref{nu_obs}. Throughout the paper, reference names are used for all the pointings. This work focuses on the newest three \textit{XMM} observations (XMMC, XMMD, XMME) and the \textit{NuSTAR} observation (taken in the form of two immediately subsequent pointings) simultaneous with XMME, i.e. parts two and three of our observing campaign described in the previous section. Part one will be presented together with an analysis of the temporal properties in a furthcoming paper.

\begin{table*}[ht]
\caption{Overview of \textit{XMM}, \textit{Suzaku} and \textit{NuSTAR} observations and their observation time, net exposure and name for further reference throughout the paper. The fluxes are given in $10^{-11}\,\text{erg}\,\text{s}^{-1}\,\text{cm}^{-2}$.}\label{nu_obs}
\begin{center}
\begin{tabular}{|c|c|c|c|c|c|c|c|}
\hline
Satellite & ObsID & Start date & Exposure & Soft flux  & Hard Flux & Reference Name & Newly analyzed here? \\
 & & & [ks] & [0.7-2.0\,keV] & [3-10\,keV] & & \\
\hline \hline \textit{XMM} & 0101040201 & 2000-07-05 & 26 & $0.59$ & $0.95$ & XMMA & N \\
 \textit{XMM} & 0605800401 & 2009-12-09 & 91 & $0.74$ &$1.2$ & XMMB & N\\
 \textit{XMM} & 0721110101 & 2013-12-19 & 42 & $0.82$ & $1.2$ &  XMMC & Y\\
 \textit{XMM} & 0721110201 & 2014-01-02 & 32 & $1.7$ & $1.8$ & XMMD & Y\\
 \textit{XMM} & 0741330101 & 2014-05-09 & 82 & $1.5$ & $1.7$ & XMME & Y\\
 \textit{Suzaku} & 702043010 & 2007-06-07 & 140 & $1.4$ & $1.8$ & SUZA & N \\	
 \textit{Suzaku} & 705063010 & 2010-05-19 & 143 & $1.6$ & $1.8$ & SUZB & N \\
 \textit{NuSTAR} & 60001130002 \& 60001130003 & 2014-05-09 & 143 & -- & $1.7$ & NU & Y \\
 \hline
\end{tabular}
\end{center}
\end{table*}

\begin{figure}[ht]
\includegraphics[width=\columnwidth]{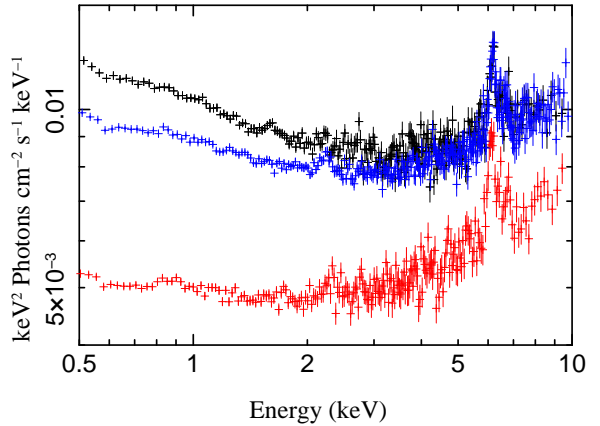}
\caption{Unfolded \textit{XMM} spectra (XMMC [red], XMMD [black], XMME [blue]) that show the variability seen in the X-ray spectrum of Fairall~9. The bump at around 2\,keV stems from the incorrect gain calibration.}\label{spectra_unfolded}
\end{figure}

\subsection{XMM-Newton}
The \textit{XMM}-\textit{Newton} satellite is equipped with two types of X-ray detectors: 
The CCD type European Photon Imaging Cameras (EPIC) and the Reflection Grating Spectrometers (RGS). The European photon imaging cameras are MOS\,1, MOS\,2, and pn. The RGS camera consists of two similar detectors, which both have high effective areas and spectral resolutions between 6 and 38\,{\AA} \citep{DenHerder2001}. The MOS cameras are aligned with the RGS detectors and have higher spatial resolution than the pn camera but a smaller effective area. 

All \textit{XMM} data were reduced with the \textit{XMM-Newton} Science Analysis System (SAS) v14.0.0. We processed the RGS data with the SAS task \texttt{rgsproc} according to the standard procedures to produce event files, spectra and response matrices.


The RGS 1 and 2 spectra of each exposure were converted to SPEX\footnote{www.sron.nl/spex} format 
through the SPEX task \texttt{trafo}. We binned the RGS spectra by a factor of 3, which provided spectral bins of about 1/3 of the FWHM and the optimal bins size avoiding oversampling (see, e.g., Fig. 3).

From the EPIC detectors we used only the pn data as the spectral modeling would be prolonged significantly by including MOS (see \S \ref{res_ccd}). After reprocessing the EPIC data using the CCF files from 2014 November, spectra were extracted using the tool \texttt{evselect}, selecting the default grade pattern (i.e. single and double events). Both source and background spectra were taken from a circular region with radius 32.5 arcsec in the case of XMMC and XMMD. For XMME a larger source region of 60 arcsec was used to ensure agreement with the simultaneous \textit{NuSTAR} observation. The source regions were centered on the source and the background regions were on the same chip near the source region. The \texttt{arfgen} and \texttt{rmfgen} tasks were used to produce the effective areas and redistribution matrices.

During the spectral modeling, energies from 0.5 to 10\,keV were considered, and the pn spectra were rebinned to a signal-to-noise of 10. We exclude the data below 0.5\,keV as here the spectra of the two EPIC cameras (pn and MOS) do not agree. As noted in \citet{Marinucci2014}; recent \textit{XMM}-EPIC-pn data are also affected by gain problems (Fig.~\ref{spectra_unfolded}), which are corrected by including a \texttt{gainshift} throughout the spectral modeling.  

\subsection{NuSTAR}
The \textit{NuSTAR} satellite \citep{Harrison2013} has two identical hard X-ray CZT focal planes, referred to as focal plane modules A (FPMA) and B (FPMB). We reduced the data from FPMA and FPMB separately and the resulting spectra are both included in the analysis. First, cleaned event files were produced using \texttt{nupipeline} v0.4.3 and \textit{NuSTAR} CALDB version 20141107. For the spectral extraction, the task \texttt{nuproducts} was used. The source region was circular with a radius of 60 arcsec, to optimize the signal-to-noise at higher energies, and was centered on the target. The background region is selected to be also circular with 60 arcsec radius and chosen close to the source region. The two spectra (from the two consecutive pointings) for each detector were co-added to produce the average spectrum.  

Similar to the \textit{XMM} spectra, the \textit{NuSTAR} spectra were also rebinned to a signal-to-noise of 7. This ensures that the spectra carry about the same weight in the fitting. The energy range considered in the spectral modeling for \textit{NuSTAR} is 3 to 80\,keV.
\section{Results}\label{results}

To investigate the nature of the flux dips and the soft excess, we first study the high-resolution RGS spectra to determine whether absorption could play a role. This analysis is then followed by broad-band modeling of the EPIC-pn and \textit{NuSTAR} spectra.

\subsection{High spectral resolution RGS data}\label{rgs_data}

Our RGS analysis focuses on the first order RGS spectra ($7-35$\,{\AA}). The second order spectra have poor statistics and are ignored in this analysis. We model the spectra with SPEX version 2.03.03. We scale the elemental abundances to the proto-Solar values of \citet{Lodders2009}, which are the default in SPEX. 
We adopt C-statistics and $1\,\sigma$ errors. At first, we model the RGS spectra of each exposure separately, where the RGS 1 and 2 spectra of the same exposure are fitted with the same spectral model.

The AGN emission within the narrow RGS energy band is described with a single, redshifted, power law absorbed by the cold interstellar medium in our Galaxy (\texttt{hot} model in SPEX with temperature fixed to 0.5 eV and column density fixed to $3.1\times10^{20}$\,cm$^{-2}$). This model is able to describe all the RGS spectra well with $\chi^2_r\sim1.2$ for 6802 degrees of freedom. One possibility is that the exceptional variability of Fairall\,9 might be due to variability of a warm absorber present in our line of sight. A warm absorber is a photoionized absorber commonly found in Seyfert galaxy X-ray spectra. The RGS is the best instrument to detect warm absorbers, and we search for a warm absorber by multiplying the power-law continuum by a photoionization \texttt{xabs} absorption model with a covering fraction of unity. The \texttt{xabs} model in SPEX calculates the transmission of a slab of material, where all ionic column densities are linked through a photoionization model. The line-of-sight velocity is a free parameter in the fit ($-5000\,\mathrm{km/s}<v<0\,\mathrm{km/s}$) because warm absorbers are often observed to be outflowing. We also allow the ionization state of the potential absorber to vary freely ($-4<\log\xi<5$). Such a component does not improve the fit significantly ($\Delta\chi^2=5$ for -3 dof), as expected, since no clear rest frame absorption lines are seen in the spectra. We estimate an upper limit on the hydrogen column density of any highly ionized gas of about $6\times10^{19}$\,cm$^{-2}$. Similarly, no rest-frame neutral absorption is detected.

\begin{figure*}[ht]
\begin{center}
\includegraphics[width=0.5\textwidth,angle=-90,bb=110 110 520 680]{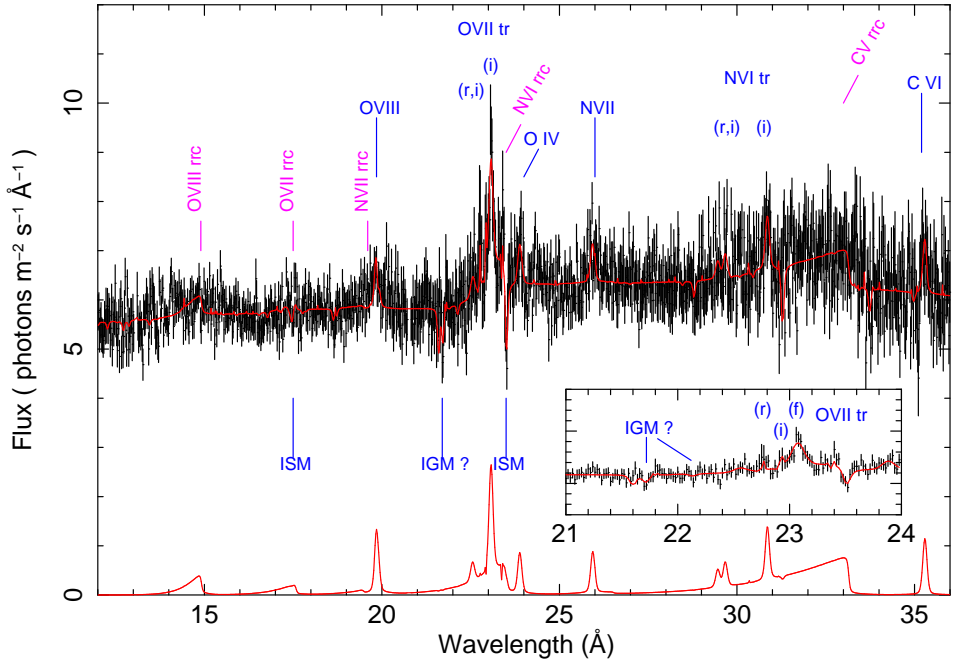}
\end{center}
\vspace{-0.3cm}
      \caption{The XMM-\textit{Newton}/RGS stacked spectrum of Fairall\,9 with the best fit model 
               adapted from the simultaneous fit of the different observations overlayed (upper red solid line). The model without the continuum power law is also shown for clarity (lower red solid line). 
               The spectrum shows evidence of the expected Galactic absorption as well as several emission lines
               and radiative recombination continua (rrc). Putative signs of absorption by the intergalactic medium are found at 21.7\,{\AA}. The emission lines are labeled and \textit{r} indicates the recommendation line, \textit{i} the intercombination line and \textit{f} the forbidden line. The small insert in the right hand corner shows a zoom-in into the wavelength region around the tentative intergalactic absorption line.}
          \label{fig:rgs}
\end{figure*}

The individual RGS spectra show evidence of narrow emission features just near the rest-frame wavelengths of the transitions of ions commonly found in high-resolution X-ray spectra of AGN: O\,VIII (19.0\,{\AA}), O\,VII triplet ($\sim$22.0\,{\AA}), N\,VII (24.8\,{\AA}), N\,VI triplet ($\sim$29.0\,{\AA}), and C\,VI (33.74\,{\AA}). These lines are apparent in the stacked spectrum in Fig.\,\ref{fig:rgs}. The fit of each spectrum improves if we model these lines with 10 Gaussians fixed to their redshifted rest frame wavelengths. The emission lines do not show any significant variability between observations, but this might be due to the low statistics.
Therefore, in order to place more constraints on the line parameters, we fit the RGS spectra simultaneously using the option \textit{sectors} in SPEX. Different observations represent different sectors, 
while RGS 1 and 2 of the same observation give one sector. The spectral models of the different observations are identical apart from the power-law normalization and slope. All the line parameters are reported in Table\,\ref{t:RGS_lines}. The 1s--2p emission lines of O\,VIII, N\,VII, C\,VI and the forbidden lines of O\,VII and N\,VII were detected at a high confidence level. On average, the emission lines have a blue shift $\Delta v=900\pm500$\,km\,s$^{-1}$ with respect to Fairall~9 and a velocity broadening $\sigma_v=1500\pm300$\,km\,s$^{-1}$.

\begin{table}[h]
\renewcommand{\tabcolsep}{1mm}
\renewcommand{\arraystretch}{1.2}
\footnotesize
\caption{\label{t:RGS_lines} Emission lines and their characteristics as derived from modeling the RGS spectra.
The line fluxes are in units of $10^{-5}$ ph s$^{-1}$ cm$^{-2}$ and the errors are given at the 68\% level ($1\sigma$).}
\begin{tabular}{ccccc}
\hline
$\lambda_0$\,(\AA)  & Ion         & Flux            & Sigma  & $V_{\rm offset}$ (km\,s$^{-1}$)\\
\hline                                                                                                                                     
18.969              & O\,VIII     & 1.5 $\pm$ 0.3   & 5.5      &        -1250 $\pm$  370       \\
21.602              & O\,VII (r)  & 1.9 $\pm$ 0.8   & 2.4      &         -430 $\pm$  380             \\
21.807              & O\,VII (i)  & 0.2 $\pm$ 0.2   & 1.0      &            0 $\pm$  790             \\
22.101              & O\,VII (f)  & 4.9 $\pm$ 0.6   & 8.5      &        -1130 $\pm$  230             \\
22.800              & O\,IV       & 0.9 $\pm$ 0.4   & 2.2      &          -70 $\pm$  340             \\
24.781              & N\,VII      & 1.7 $\pm$ 0.3   & 5.1      &         -260 $\pm$  680             \\
28.787              & N\,VI (r)   & 0.3 $\pm$ 0.3   & 1.0      &        -7300 $\pm$ 5300             \\
29.084              & N\,VI (i)   & 0.8 $\pm$ 0.4   & 2.2      &        -7900 $\pm$ 4000             \\
29.534              & N\,VI (f)   & 2.4 $\pm$ 0.4   & 5.5      &         -930 $\pm$  280             \\
33.736              & C\,VI       & 3.3 $\pm$ 0.6   & 5.4      &         -230 $\pm$  380             \\
\hline
14.23               & O\,IX$\rightarrow$O\,VIII RRC   &  2.3 $\pm$ 0.7   &  3.8   & $\equiv0$   \\
16.78               & O\,VIII$\rightarrow$O\,VII  RRC &  1.5 $\pm$ 1.0   &  1.5   & $\equiv0$   \\
18.59               & N\,VIII$\rightarrow$N\,VII  RRC &  0.6 $\pm$ 0.2   &  2.6   & $\equiv0$   \\
22.46               & N\,VII$\rightarrow$N\,VI   RRC  &  6.3 $\pm$ 1.8   &  3.5   & $\equiv0$   \\
25.30               & C\,VII$\rightarrow$C\,VI   RRC  &  0.6 $\pm$ 0.6   &  1.0   & $\equiv0$   \\
31.63               & C\,VI$\rightarrow$C\,V   RRC    & 17.5 $\pm$ 3.5   &  5.0   & $\equiv0$   \\
\hline
\end{tabular}
\end{table}

The stacked spectrum also shows evidence of some broad emission-like feature found just at the rest-frame
wavelengths of some relevant radiative recombination continua (RRCs). 
They can be modeled with the \texttt{rrc} model, which calculates radiative recombination continua for photoionized plasmas (see SPEX manual for details). The recombining plasma has an average low temperature $kT_e\sim$10\,eV, suggesting photoionization. The fluxes of the RRC produced by each ion are reported in Table\,\ref{t:RGS_lines}. We caution that the C\,VI$\rightarrow$C\,V RRC at 31.63\,{\AA} may be affected by large systematic uncertainties due to background subtraction.

There is no evidence of neutral or ionized reflection in the rest-frame of Fairall~9. However, in the brightest spectra, there is weak evidence of absorption at 21.7\,{\AA}, 
which also appears in the stacked spectrum in Fig.\,\ref{fig:rgs}.
In the rest-frame of Fairall\,9 this would correspond to 20.7\,{\AA},
where no strong warm absorber feature is expected. The closest strong transition is the O\,VII resonance line at 21.6\,{\AA}. 

\subsection{Broad-band data}\label{res_ccd}

Our broad-band spectral modeling was performed using Isis v1.6.2-27. The errors quoted in this section are at 90\,\% confidence.

We built our broad-band spectral model based on previous detailed spectral studies. These studies revealed that the spectrum of Fairall~9 is a superposition of the following components: a continuum (usually a power law), a distant neutral reflector, relativistically blurred ionized reflection, highly ionized photoionized emission and possibly an additional soft X-ray excess component \citep{Emmanoulopoulos2011,Patrick2011,Lohfink2012a,Walton2012}. All components are modified by Galactic absorption with a column of $3.15\times 10^{20}\,\text{cm}^{-2}$. The lack of evidence for ionized absorption or excess neutral absorption in the RGS data indicates that one (or more) of the these components must be responsible for the variability we observe (Fig.~\ref{spectra_unfolded}). To determine which one(s), the three \textit{XMM} observations (XMMC, XMMD, XMME) were fitted jointly in a multi-epoch fit, in which the \textit{NuSTAR} (NU) is also included.

\begin{figure*}[ht]
\begin{center}
\includegraphics[width=0.75\textwidth]{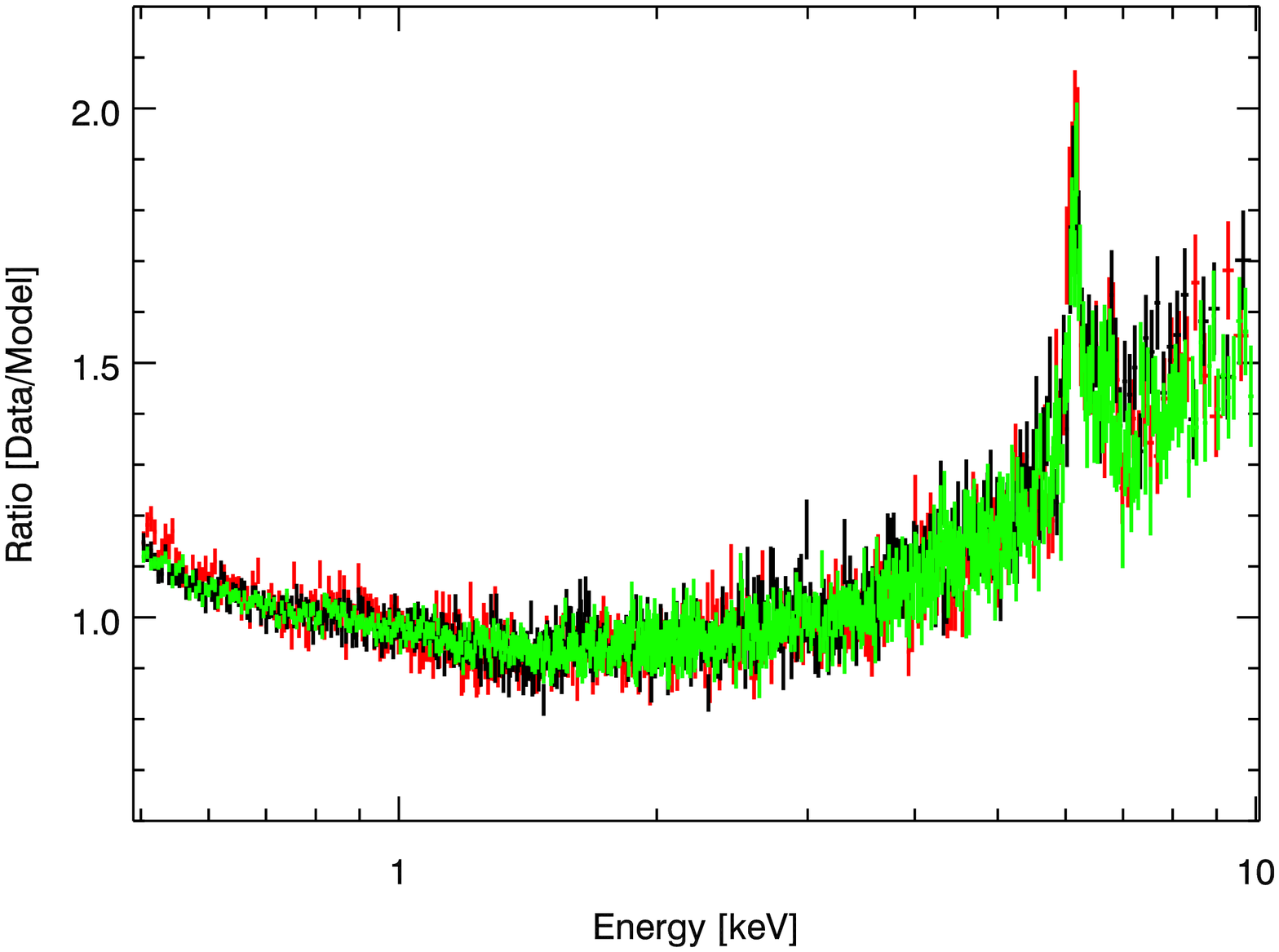}
\caption{Ratio to a simple absorbed power law for the newest \textit{XMM} observations: XMMC [black], XMMD [red], and XMME [green]. The ratio residuals are the same for all three pointings, indicating that the fractional contribution of the different model components at any given energy is constant with time. The data have been rebinned slighty for plotting clarity.}\label{resid_powerlaw}
\end{center}
\end{figure*}

\begin{figure}[ht]
\includegraphics[width=\columnwidth]{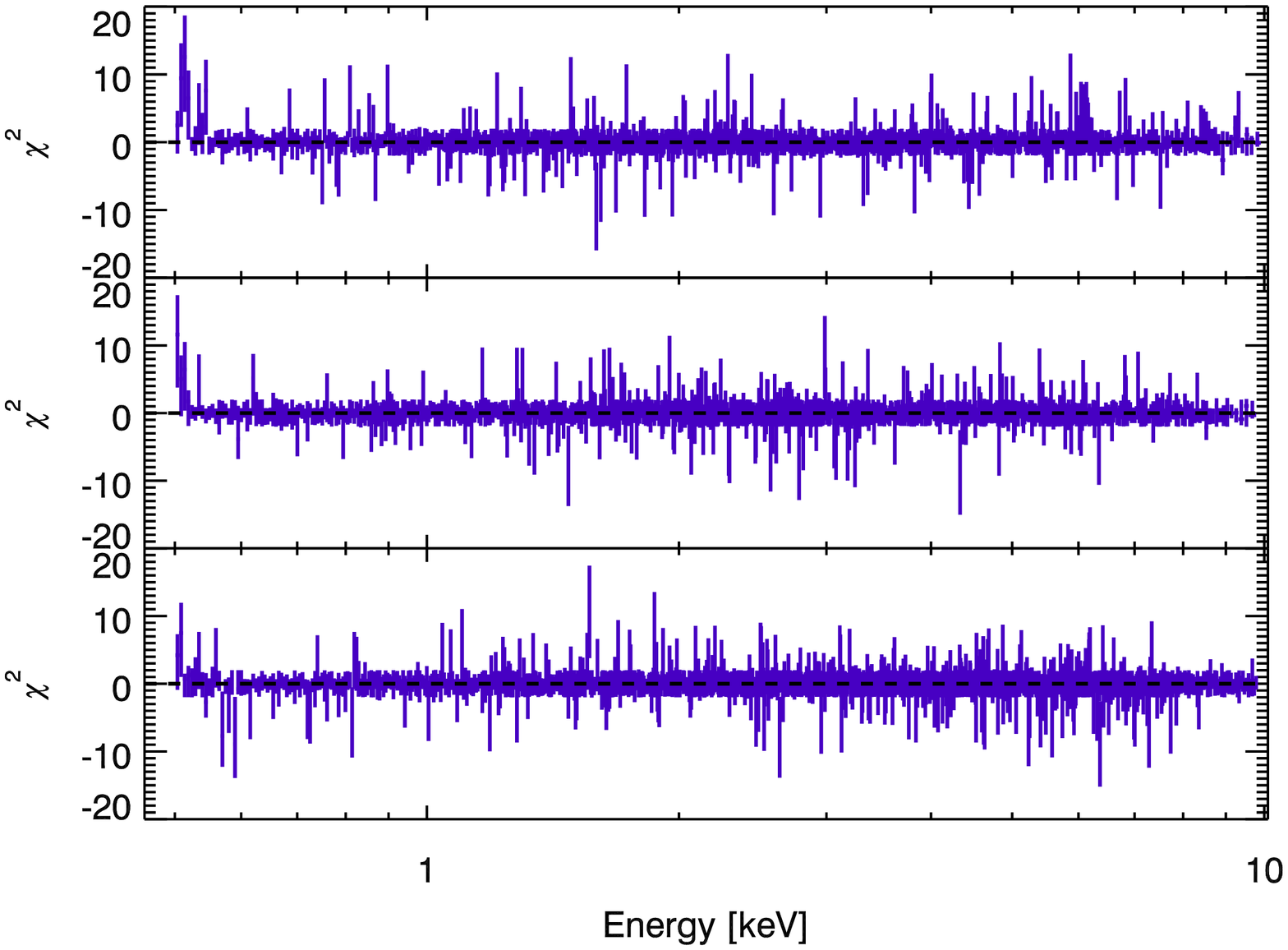}
\caption{Residuals to a simple power law times the (weighted) average residuals of the three \textit{XMM} observations to a simple power law. The panels correspond to: XMMC [top panel], XMMD [middle panel], and XMME [bottom panel]. The residuals at the very softest energies originate from narrow emission lines (\S \ref{rgs_data}) that do not scale with the power law.}\label{resid_powerlaw_refl}
\end{figure}

First, however, to confirm that the spectral shape has not significantly changed from previous observations, we fit an absorbed power law (both photon index and norm are fitted) to the full band (0.5-10\,keV) of each of the \textit{XMM} spectra. The \textit{NuSTAR} spectra are ignored for this initial exploration. The best fit residuals are shown in Figure~\ref{resid_powerlaw}. It is clear that the fractional residuals (data/model) with respect to this power law model are remarkably similar for all three observations. The presence of reflection is also apparent from the strong iron line in the spectrum. Additionally, a strong soft X-ray excess is visible below 2\,keV. To assess more rigorously whether the fractional residuals to a simple power law are really constant, we calculate a weighted average of the data/model residuals. These residuals are converted into a multiplicative xspec table model, which we fit multiplied with a variable power law to the individual spectra. We find that this provides a very good description of the \textit{XMM} data (Fig.~\ref{resid_powerlaw_refl}). The only noteable deviations are present at soft X-ray energies. These residuals, on or below the resolution limit of EPIC-pn, are at the 5\,\% level and are most likely not physical. A combination of small inaccuracies in the gain correction applied to the spectra\footnote{We assumed the gain correction to be linear with energy, but this is only a simple approximation.} and in the background subtraction can easily lead to a 5\,\% deviation. These simple studies of the new spectra suggest the presence of the same spectral components as previously observed, as the data/model ratios look similar to those shown in Figure~2 of \citet{Lohfink2012a}. No new features are detected that would indicate the appearance of new spectral components. However, the fact that at each energy the fractional flux corresponding to continuum (power law), cold and probably ionized reflection is constant in the \textit{XMM}-band is new information that we can utilize when performing the detailed modeling of these spectra, as will be described in more detail below. 

Based on the finding that the spectra have the same components as observed previously, we model the spectra in the joint analysis as follows: the continuum is assumed to be a cut-off power law with the normalization and slope allowed to vary for each observation as they seem to be the cause of most of the variability (Fig.~\ref{resid_powerlaw}). The high-energy cut-off is assumed to be constant throughout the observations; although some change in the cut-off is expected \citep{Fabian2015}, it is most likely to be too small to be measurable. We note here that inclusion of a cut-off is important as it has an impact on the reflection spectrum at low energies even if its value is not in the observed band \citep{Garcia2015}. We model the cold reflector using the \texttt{pexmon} model \citep{Nandra2007} with the reflection fraction frozen to -1, putting the model in a reflection-only mode. The strength of the reflection is now controlled by the model normalization. The irradiating cut-off power law is presumed to be the primary continuum, setting the photon indexes and high energy cut-off to be the same for both. The inclination parameter of the model cannot be determined from the data itself, and we therefore tie it to that of the ionized reflector. This assumes that the cold reflection originates from the outer parts of the accretion disk, which, as we will see later in Section \ref{discuss}, is a reasonable assumption. The blurred ionized reflection, which we assume accounts for the soft excess as well, is described by the model \texttt{relxill} (version 0.2g) \citep{Garcia2014} which is also put into its reflection-only mode. Again, the primary continuum is presumed to be responsible for the irradation and the photon indexes and high energy cut-off of the reflection are tied to that of the continuum. The emissivity profile is a broken power law, with the index free to vary for radii smaller than $15\,R_\text{g}$ and fixed to 3 for larger radii. In addition to the cold and ionized reflection, the highly ionized photoionized emission is modeled with an XSTAR grid computed from the model \texttt{photemis}. As the normalization and column density of the \texttt{photemis} model are degenerate for high ionization states, we fix the column to $10^{22}\,\mathrm{cm}^{-2}$. The photoionization most likely occurs on large scales, making changes to the ionization state within half a year unlikely and we therefore tie the ionization parameter of the \texttt{photemis} component for the three observations.\footnote{This assumption has no impact on the results obtained here.} Furthermore, cross calibration constants are included for the \textit{NuSTAR} observation, which are defined with respect to the simultaneous XMME observation. Finally, everything is altered by the absorption from the ISM in our Galaxy using the model \texttt{TBnew} \footnote{http://pulsar.sternwarte.uni-erlangen.de/wilms/research/tbabs/} with abundances set to \texttt{wilm} and cross sections to \texttt{vern}.

No changes are expected to the black hole spin, inclination of the accretion disk or iron abundance on the timescales of our observations (Fig~\ref{light_f9}). We can therefore use this information and tie these fit parameters of the different observations together in the joint fit. 

\begin{figure}[ht]
\includegraphics[width=\columnwidth]{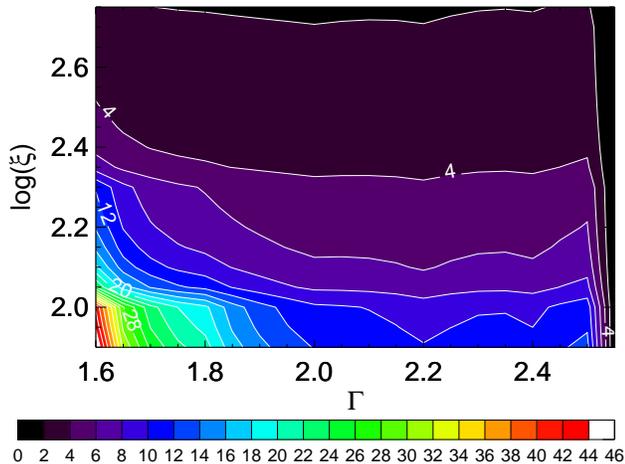}
\caption{Ratio of power law flux and \texttt{relxill} flux at 1\,keV for different values of $\Gamma$ and $\log(\xi)$. For higher ionization states the two flux values approach each other. A similar trend is also seen with increasing photon index.}\label{sim_relxill}
\end{figure}

The fractional model contributions of the non-continuum components at a given energy remain constant, as we saw earlier. We determine these fractions from one spectrum and then apply them to the others. In practical terms we want to determine and then enforce a model flux ratio between different model components at a given energy. The power law continuum and the \texttt{pexmon} model are already normalized to the photon flux at 1\,keV. We therefore use 1\,keV as our reference point. The flux contribution of the \texttt{relxill} model is not normalized to 1\,keV and the model flux at 1\,keV depends upon most of its parameters, including the photon index ($\Gamma$), the black hole spin ($a$), the inclination of the accretion disk ($i$), and the ionization state ($\log(\xi)$). To investigate the impact of the individual parameters on the flux ratio between the power law continuum and the \texttt{relxill} model, we calculate the model flux of \texttt{relxill} at 1\,keV for the normalization set to 1, varying photon indexes and ionization states in steps of 0.05. While $a$ and $i$ also influence the flux normalization, they remain constant with time, and only cause a near-constant offset, which does not affect our modeling. We therefore fix $a$ to $0.998$ and $i$ to $30$\,deg. During the whole process, the \texttt{relxill} model is switched to its reflection-only mode by setting the reflection fraction to -1, similar to what is done during the actual modeling. The resulting flux ratios (where $f$=(power law flux)/(reflection flux)) with respect to $\Gamma$ and $\log(\xi)$ are shown in Figure~\ref{sim_relxill}. It is apparent that higher ionization states and increasing photon indexes return flux ratios closer to one. This can be explained by the reflection spectrum approaching a power law-like shape. 

To use this information in the modeling we fit the parameter space of interest , i.e. $\log(\xi)=1.90-2.75$, and $\Gamma=1.60-2.55$, with a second order polynomial:
\begin{equation}\label{eqn}
f=a\times\Gamma^2+b\times\Gamma+c+d\times\log(\xi)^2+e\times\log(\xi),
\end{equation}where $a$, $b$, $c$, $d$, and $e$ are free parameters. We determine the parameters to $a=21.63$, $b=-95.51$, $c=301.81$, $d=28.97$ and $e=-150.85$. With this function we have a prediction of the model `intrinsic' flux ratios due the definition of the normalizations. This factor is taken into account in the spectral modeling and the true flux ratio kept constant. This is done by keeping $(N_\mathrm{pow}/N_\mathrm{refl})\times 1/f$ constant. The spectrum XMMC is used to determine the flux contribution of the different components at 1\,keV and the same flux ratios are enforced for the spectra XMMD and XMME/NU. This means that for XMMD and XMME/NU, the power law normalization is the only normalization free to vary. The normalization of \texttt{relxill} is derived using Equation 1. The normalization of \texttt{pexmon} can be derived simply from the ratio of normalizations for XMMC.

With this model set-up we obtain a reasonable description of the data ($\chi^2=4203.8$ for 3708 dof). As in the last section (\S \ref{rgs_data}) where we identified several emission lines in the soft X-ray band in the RGS data (consistent with a low ionization photoionized component), one would expect these lines to also be present in the pn spectra. We therefore include these lines in the joint broad band fit. As the lines are rather weak and hard to constrain with pn data, we keep their energies and normalizations fixed to the average best fit values from the RGS analysis. The inclusion of the lines improves $\chi^2$ by 6.3, showing that the lines are also consistent with the pn data. Besides the improvement in $\chi^2$, the inclusion of the lines does not bring about notable changes to the other fit parameters (Table~\ref{wlines}), as expected for such weak lines. We find that the primary continuum  is softer as the source is brighter, i.e. $\Gamma=2.15\pm0.01$ for XMMC and 2.33$\pm+0.01$ for XMMD. Despite those changes of the primary continuum, the ionization state of the source remains almost constant (Table~\ref{wlines}). From the joint fit, we also get very good constraints on the black hole spin (0.973$\pm0.003$) and accretion disk inclination ($<11$\,degrees), indicating an almost face-on disk.  

\begin{table*}[ht]
\caption{Spectral parameters for a model with a power law, a cold reflector, blurred ionized reflection, highly ionized photoionized emission and low energy emission lines fitted to the newly analyzed \textit{XMM} and \textit{NuSTAR} data. The parameters marked with $d$ have been derived using Equation\,1. The cut-off powerlaw and \texttt{pexmon} normalization is the photon flux in photons\,keV$^{-1}$\,cm$^{-2}$\,s$^{-1}$ at 1\,keV, in case of \texttt{pexmon} reflection is not taken into account in the flux. Bold parameter values indicate that these parameters were determined jointly from all datasets by tieing the parameter during the fitting.}\label{wlines}
\begin{center}
\begin{tabular}{c|c|c|c|c}
& & XMMC & XMMD & XMME/NU \\
\hline \hline Cold Reflection & $\Gamma$ & 2.15$\pm0.01$  & 2.33$\pm0.01$ & 2.20$\pm0.01$ \\
 & Fe/Solar & \multicolumn{3}{c}{\textbf{0.83}$_{-0.04}^{+0.06}$}  \\
 & $N_\text{pex} [10^{-3}]$ & 6.3$_{-0.3}^{+0.4}$ & 13.2$^d$ & 11.3$^d$  \\
\hline Continuum & $N_\text{pow} [10^{-3}]$ & 4.27$_{-0.06}^{+0.10}$ & 8.92$_{-0.12}^{+0.11}$ & 7.65$_{-0.13}^{+0.19}$  \\
& $E_\text{cut}$ [keV] & \multicolumn{3}{c}{\textbf{$>$933}} \\
\hline Ionized Reflection & $N_\text{rel} [10^{-3}]$ & 4.10$_{-0.27}^{+0.25}$ & 8.80$^d$ & 3.77$^d$ \\
 & $q$ & 7.0$\pm0.3$ & 7.0$\pm0.3$ & 6.5$_{-0.2}^{+0.3}$ \\
 & $a$ & \multicolumn{3}{c}{\textbf{0.973}$\pm0.003$} \\
 & $i$ & \multicolumn{3}{c}{\textbf{$<$11}} \\
 & $\log \xi$ & 2.17$\pm0.02$ & 2.17$\pm0.02$ & 2.70$\pm0.01$ \\ 
\hline Photoemission & $N_\text{phot} [10^{-3}]$ & $<4.7$ & $4.3_{-3.8}^{+6.3}$ & 5.2$_{-2.4}^{+4.8}$ \\
& $\log \xi$ & \multicolumn{3}{c}{\textbf{4.08}$_{-0.33}^{+0.28}$}  \\   
\hline Cross Calibration &  $c_\text{fa}$ & - & - & 1.02$\pm0.01$ \\
  & $c_\text{fb}$ & - & - & 1.06$\pm0.01$ \\
\hline \hline  & $\chi^2$ & \multicolumn{3}{c}{4197.5} \\
  & dof & \multicolumn{3}{c}{3705}  \\
\end{tabular}
\end{center}
\end{table*}

\begin{figure*}[ht]
\begin{center}
\includegraphics[width=1.75\columnwidth]{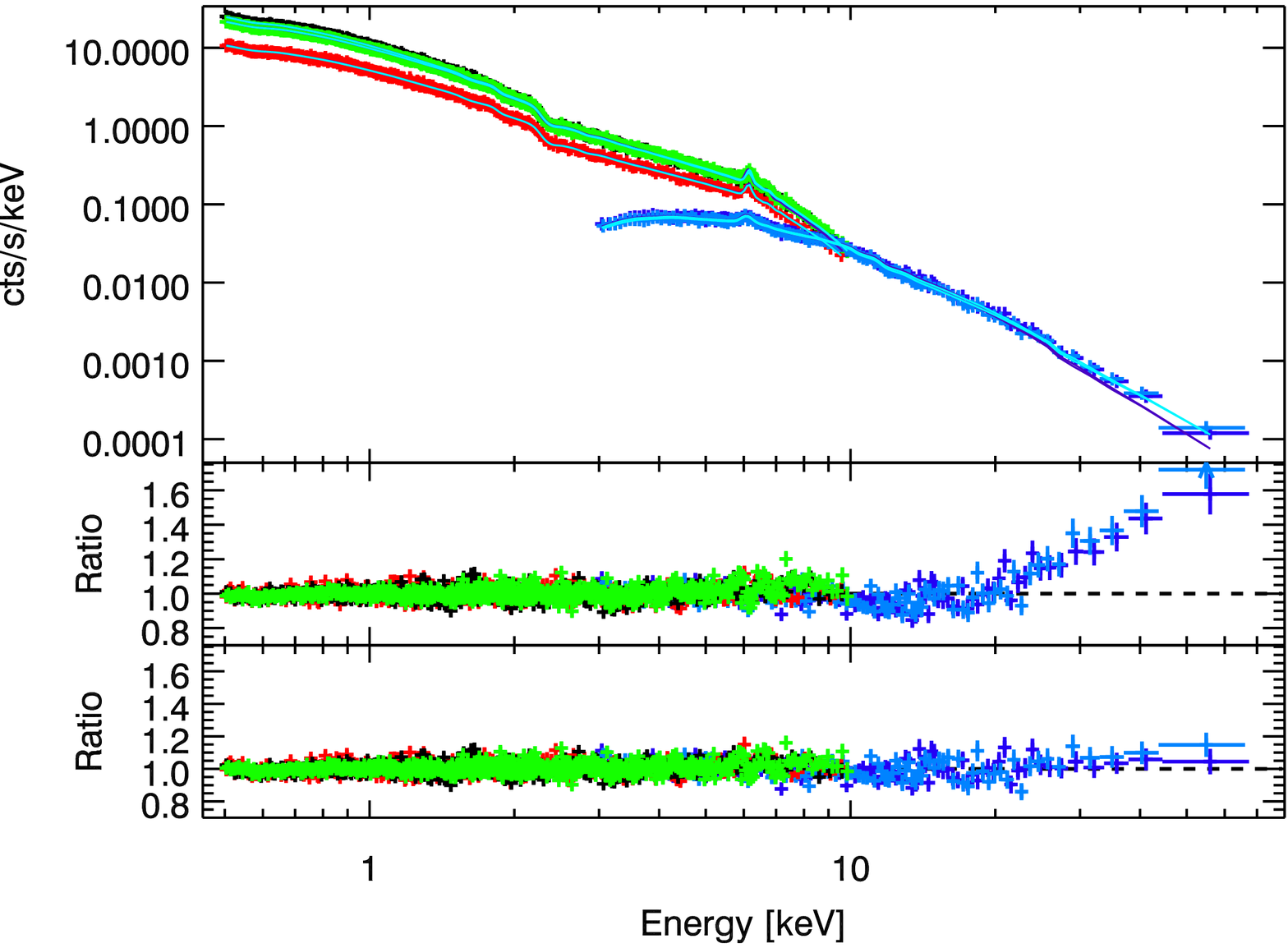}
\end{center}
\caption{ Data and ratio residuals to a broad-band fit with a soft excess modeled by a) only one blurred ionized reflector [middle panel] and b) a blurred ionized reflector and an additional soft power law [bottom panel]. The former model fails to account for the observed flux at hard X-rays. The data shown are XMMC [black], XMMD [red], XMME [green], and NU [blue]. The data, model and residuals have been rebinned slightly for clarity.}\label{residuals}
\end{figure*}

Despite the seemingly acceptable $\chi^2$ value, the residuals from the fit reveal that the model fails to properly describe the \textit{NuSTAR} spectrum at energies above 20\,keV (Fig.~\ref{residuals}, middle panel). The model clearly underpredicts the observed hard X-ray flux, indicating that the model components included in this fit cannot describe the data satisfactorily. The acceptable $\chi^2$ is driven by the large number of bins below 10\,keV relative to above 10\,keV. 

\subsection{The soft X-ray excess}\label{soft}

In the previous section we found that a spectral model including continuum emission, distant reflection and blurred ionized reflection from a homogeneously ionized inner accretion disk fails to describe the data well, leaving large residuals in the \textit{NuSTAR} band. One possible explanation of this excess could be that the inner disk is not all at the same ionization state, but instead the ionization state decreases with radius. The total blurred ionized reflection spectrum is then a superposition of the spectra at different radii. To explore this scenario we replace the \texttt{relxill} component in the model with the model \texttt{relxill\_ion}. \texttt{relxill\_ion} calculates the relativistically smeared reflection spectrum in a manner similar to \texttt{relxill}, but for a number of zones, where the number of the zones is provided by the user. The ionization states of the individual zones are defined via an ionization gradient. For our modeling we select five zones with an ionization gradient of $r^{-3}$. While selecting more zones would yield more exact results, the computational time required for the spectral modeling and error calculation would become prohibitive ($\sim$ several weeks) for our multi-epoch broad-band fit. Note that the flux ratios cannot be enforced for the ionized reflection anymore as Eqn.\ref{eqn} is not valid anymore. Considering an ionization gradient in the inner disk leads to no improvement in $\chi^2$. The fit also still fails to match the data at high energies and we therefore do not consider it further. 

\begin{figure}[h]
\includegraphics[width=\columnwidth]{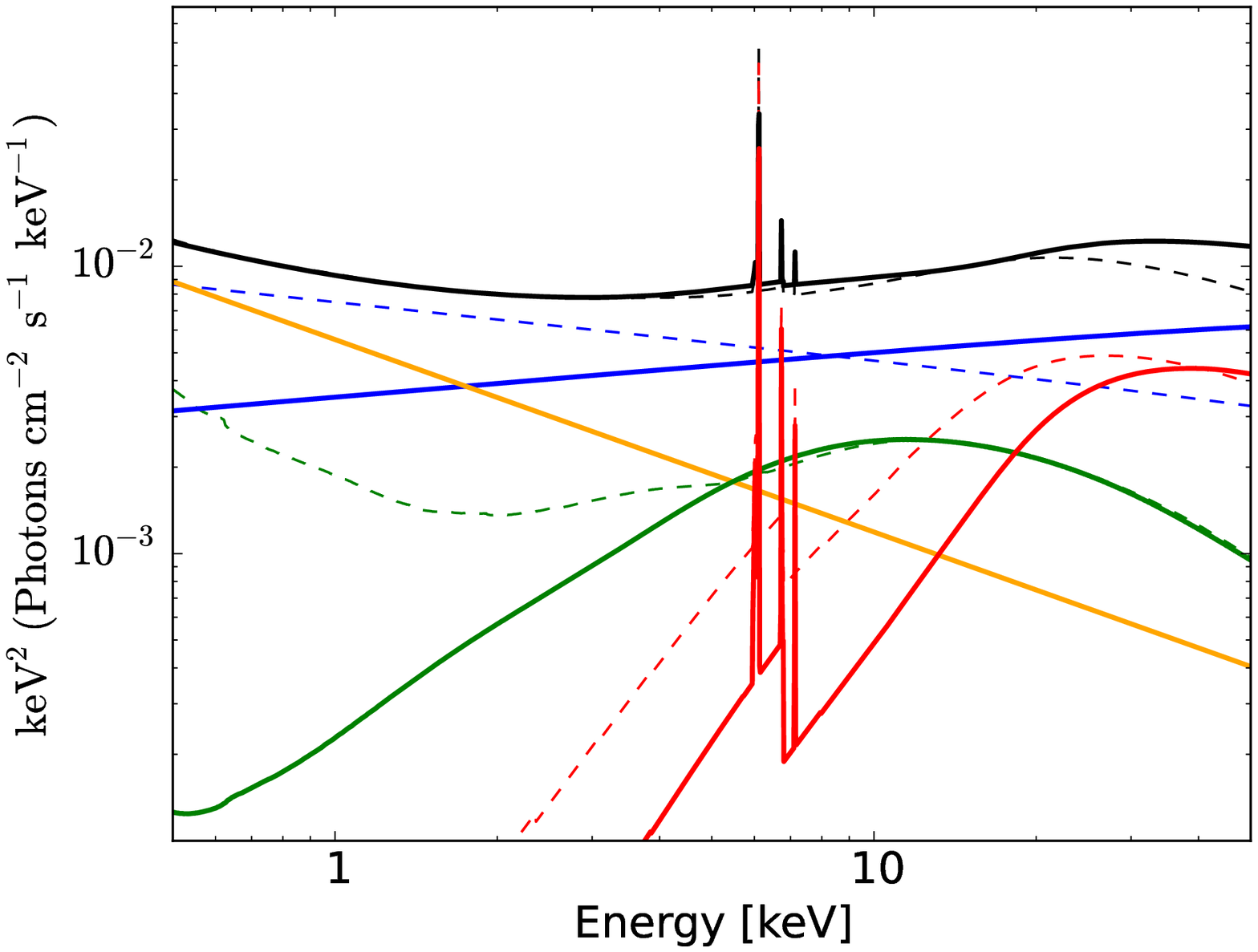}
\caption{The best fit unfolded spectral models for the XMME/NU data. The colored lines correspond to: the total model [black], primary power law [blue], neutral reflection [red], blurred ionized reflection [green], and soft power law [orange]. The components drawn with solid lines are those from a model with an additional soft power law and the dashed lines are those without. The total model fluxes are almost identical below 20\,keV; only at higher energies the model with the soft power law predicts more flux. For clarity only the main broad-band spectral components are shown. The Galactic absorption and emission lines (both low and high ionization) have been omitted.}\label{model_decomp}
\end{figure}

Another possibility is to test whether the inclusion of an additional soft excess component brings an improvement to the high energy flux prediction. We showed in a previous paper \citep{Lohfink2012a} that the modeling of the soft excess does affect the predicted hard X-ray flux in the case of Fairall 9. To study the effect on the new data, we include an additional \texttt{power law} component in our model. To ensure that the fractional model flux contribution stays constant with respect to the power law, we assume that the difference between the primary photon index and the secondary soft power law photon index stays constant for all the three observations, i.e. $\Gamma-\Gamma_\text{soft}=\text{constant}$. The ratio of the power law normalizations remains constant for the different datasets. The inclusion of the power law leads to a significant improvement to the fit statistic ($\Delta\chi^2=249.3$) with respect to the simple reflection model and models the hard X-ray data well (Figure~\ref{residuals}, bottom panel). The parameters of this fit are very similar to the simple reflector fit, except that part of the continuum flux is in the additional power law, and the photon index of the primary power law is smaller (Table~\ref{wpow}). This is also visible in a comparison of the model decomposition of the XMME/NU spectra for the case with and without an additional soft power law (Figure~\ref{model_decomp}). Despite the much improved description at hard X-rays some residuals remain, these can be quantified by comparing the average weighted $\chi^2$ residuals below and above 20 keV. We find the average residuals below 20\,keV to be $0.30\pm0.02$ and above 20\,keV to be $0.42\pm0.08$. This minor hard excess could simply be due to the physical limitations of the models used in describing the data, as no other hard X-ray component would be expected for a radio-quiet Seyfert. 
In a final effort to improve the fit further, we modify the additional soft power law to cut-off at higher energies. However, we find that the inclusion of a cut-off does not yield any further improvement to the fit statistic.   

\begin{table*}[ht]
\caption{Spectral Parameters for a model with a power law, a cold reflector, blurred ionized reflection, highly ionized photoionized emission, low energy emission lines and additional soft power law fitted to the newly analyzed \textit{XMM} and \textit{NuSTAR} data. The parameters marked with $d$ have been derived using Equation\,1. The cut-off powerlaw, soft power law and \texttt{pexmon} normalization is the photon flux in photons\,keV$^{-1}$\,cm$^{-2}$\,s$^{-1}$ at 1\,keV, in case of \texttt{pexmon} reflection is not taken into account in the flux. Bold parameter values indicate that these parameters were determined jointly from all datasets by tieing the parameter during the fitting.}\label{wpow}
\begin{center}
\begin{tabular}{c|c|c|c|c}
& & XMMC & XMMD & XMME/NU \\
\hline \hline Cold Reflection & $\Gamma$ & 1.73$_{-0.04}^{+0.03}$  & 1.92$\pm0.03$ & 1.84$_{-0.04}^{+0.02}$ \\
 & Fe/Solar & \multicolumn{3}{c}{\textbf{1.9}$_{-0.2}^{+0.3}$}  \\
 & $N_\text{pex} [10^{-3}]$ & 1.4$_{-0.2}^{+0.2}$ & 2.9$^d$ & 2.5$^d$  \\
\hline Continuum & $N_\text{pow} [10^{-3}]$ & 1.91$_{-0.14}^{+0.24}$ & 3.96$_{-0.30}^{+0.39}$ & 3.51$_{-0.22}^{+0.16}$ \\
& $E_\text{cut}$ [keV] & \multicolumn{3}{c}{\textbf{784$_{-271}^{+162}$}} \\
\hline Ionized Reflection & $N_\text{rel} [10^{-3}]$ & 1.28$_{-0.32}^{+0.19}$ & 2.56$^d$ & 1.07$^d$ \\
 & $q$ & 7.1$_{-0.6}^{+1.0}$ & 7.5$_{-1.2}^{+1.1}$ & $>9.9$ \\
 & $a$ & \multicolumn{3}{c}{\textbf{$>$0.997}} \\
 & $i$ & \multicolumn{3}{c}{\textbf{$<$11}} \\
 & $\log \xi$ & 2.30$_{-0.11}^{+0.04}$ & 2.06$_{-0.09}^{+0.08}$ & 1.85$_{-0.09}^{+0.10}$ \\ 
\hline Soft Excess & $\Gamma$ & 2.56$_{-0.02}^{+0.03}$ & 2.76$^d$ & $2.67^d$ \\
 & $N_\text{soft} [10^{-3}]$ & 3.02$_{-0.10}^{+0.10}$ & 6.26$^d$ & 5.56$^d$ \\ 
\hline Photoemission & $N_\text{phot} [10^{-3}]$ & $0.76_{-0.45}^{+0.71}$ & 1.24$_{-0.81}^{+1.11}$ & 1.16$_{-0.41}^{+0.52}$ \\
& $\log \xi$ & \multicolumn{3}{c}{\textbf{3.47$_{-0.02}^{+0.09}$}}  \\   
\hline Cross Calibration &  $c_\text{fa}$ & - & - & 1.02$\pm0.01$ \\
  & $c_\text{fb}$ & - & - & 1.06$\pm0.02$ \\
\hline \hline  & $\chi^2$ & \multicolumn{3}{c}{3948.2} \\
  & dof & \multicolumn{3}{c}{3706}  \\
\end{tabular}
\end{center}
\end{table*}

\section{Agreement with previous observations}\label{previous}
\begin{figure}[ht]
\includegraphics[width=1.0\columnwidth]{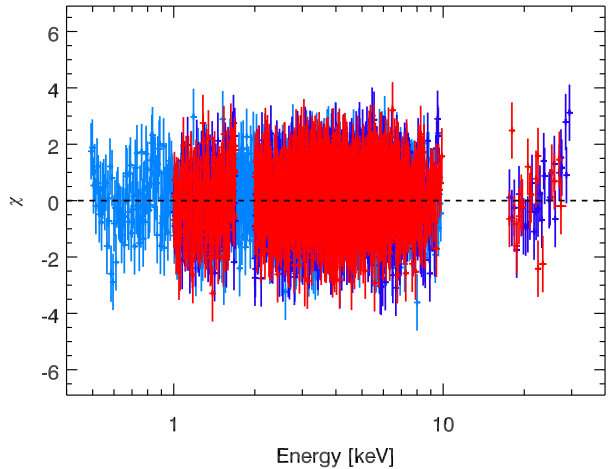}
\caption{Residuals of SUZA (blue), SUZB (red), XMMB (light blue) for a similar model to the best fit model to the one described in Section~\ref{res_ccd} found to fit the new \textit{XMM} and \textit{NuSTAR} data. The residuals below 1\,keV are consistent with the calibration uncertainity of \textit{XMM} EPIC-pn.}\label{residuals_previous}
\end{figure}

The results presented thus far are based on observations taken over a period of less than one year (Table~\ref{nu_obs}), while the observations considered in previous works spanned a period from 2000 to 2010. Changes to the geometry of the source are plausible over a several year period. These changes might cause the flux ratios to the continuum to vary. Further, we also note the different sensitivities of the different satellites leads to a different average power law fitted to even the same band, which would affect the data/model residual plots as well. Therefore, instead of trying to make a comparable plot for the old data we simply try to fit the SUZA, SUZB, and XMMB observations\footnote{The spectra used are those presented in \citet{Lohfink2012a} and a description of the \textit{Suzaku} data reduction can be found in this previous paper.} jointly in a similar fashion to the XMMC, XMMD, and XMME/NU observations to test whether the data have similar properties. For ease of fitting, we combine the two front-illuminated \textit{Suzaku} XIS detectors. We include the power law soft excess in our model, as it was already seen as a possibility for the soft excess when the same observations were analyzed in \citet{Lohfink2012a}. However it was already noted then that the soft X-ray energies of \textit{Suzaku}-XIS might not be very reliable because of contamination on the detector. To avoid this affecting our general conclusion about the spectra, we exclude all \textit{Suzaku} data below 1\,keV for the spectral modeling. The high energy cut-off is kept fixed at 1000\,keV, as it cannot be determined from \textit{Suzaku}-PIN. We find that a model similar to that described in Section \ref{soft} describes the old data reasonably well ($\chi^2_\mathrm{red}=1.08$ for 5386 dof), see Fig.~\ref{residuals_previous}. In particular such a fit returns very similar best fit parameters ($a>0.995$,$i<14\,\mathrm{deg}$) to the fit to the newer data. The spin and inclination measurements are in agreement with those found in Section~\ref{soft}. This implies that the source behavior remained mostly unchanged over the last 7 years. 

\section{Discussion}\label{discuss}
\subsection{Summary}
In this paper we have studied three \textit{XMM} observations of Fairall~9 taken in 2013/2014 and one \textit{NuSTAR} observation in the same time period.

Our main findings are:
\begin{itemize}
\item We are able to confirm the absence of excess ionized or neutral absorption in the soft X-ray band. We detect emission lines in the soft band, as well as a single absorption line at 21.7\,\AA\, that cannot be associated with Fairall~9.
\item We discover fractional residuals to a simple power law in the \textit{XMM} band, which remain constant even as the source flux changes significantly. This implies a constant fractional contribution from the different model components. 
\item We clearly detect blurred ionized reflection in the spectrum and are able to constrain the accretion disk parameters.
\item A soft excess additional to that due to blurred ionized reflection, described by an additional power law  component is present.   
\end{itemize}

\subsection{The detection of imprints from the intergalactic medium in the RGS spectra}
In section \ref{rgs_data} we noted that there is a noticeable absorption line at 21.7\,{\AA}, which agrees most closely with the O\,VII resonance line at 21.6\,{\AA}. The hot phase of the interstellar medium is known to produce prominent O\,VII absorption lines \citep[see, e.g.,][and references therein]{Pinto2013}.
However, assuming the line is produced by O\,VII absorption, it would have a velocity shift of $1400\pm200$\,km\,s$^{-1}$, which is beyond any velocity range in the Galaxy, including high velocity clouds ($|v|<400$\,km\,s$^{-1}$), or uncertainties due to errors in the RGS absolute wavelength scale (typically 0.004\,{\AA} or $\sim100$\,km\,s$^{-1}$). The most likely and appealing origin of this feature is therefore the intergalactic medium lying 
in the line of sight towards Fairall\,9.

Assuming that the absorption feature detected at $2.8\,\sigma$ at 21.7\,{\AA} is produced by O\,VII,
this would correspond to $N_{\rm OVII}=(1.4\pm0.5)\times10^{16}$\,cm$^{-2}$ at $z=0.0047\pm0.0008$.
In general, the best indicator for IGM absorption is the redshifted H\,I Ly\,$\alpha$ absorption line
(1215.67\,{\AA} at rest), which is commonly found in FUV spectra of AGN,
including Fairall\,9 \citep[see, e.g.,][]{Danforth2006}.
The redshift of our putative O\,VII feature is consistent with the H\,I absorption line
redshifted to 1221\,{\AA} clearly seen in the H\,I Ly\,$\alpha$ profile of Fairall\,9
\citep[see][Fig.\,10]{Richter2013}. The 21.7\,{\AA} line is therefore most likely of intergalactic origin. 

\subsection{The absence of absorption}
\begin{figure}
\begin{center}
\includegraphics[width=0.75\columnwidth,angle=90]{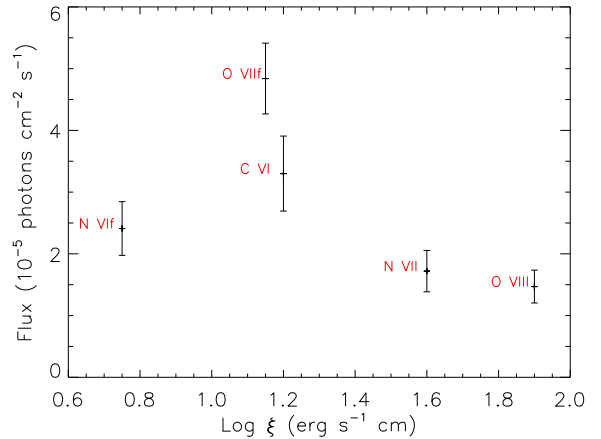}
\end{center}
      \caption{Flux versus peak ionization parameter for the lines detected with high confidence.}
          \label{fig:xi_vs_flux}
\end{figure}

From the RGS spectra, we confirm the absence of any significant absorption other than Galactic, as noted previously by \citet{Reynolds1997} and \citet{Emmanoulopoulos2011}. In fact, we see a set of low energy emission lines including those consistent with O VII, O VIII, and N VII. Plotting the flux of these lines against the ionization state (Figure~\ref{fig:xi_vs_flux}) indicates that the ionization state of most of the gas is around $\log \xi=1.1\,\text{erg}\,\text{s}^{-1}\,\text{cm}$. Such low ionization emission lines are consistent with a warm absorber out of our line of sight. Within the context of the unified AGN model, this agrees well with our finding that the inclination of the inner accretion disk, as measured from the multi-epoch spectra, is very low and the accretion disk is almost face-on. Previous measurements of the inclination angle have found it to be more intermediate \citep{Schmoll2009,Walton2012,Lohfink2012a}. This discrepancy could could be due to that earlier analyses have used an angle averaged reflection model, which the analysis presented here uses the new angle-resolved \texttt{relxill} model \citep{Garcia2014}. For very low inclinations and intermediate ionization states such as found here, the differences between an angle-averaged and angle-resolved model are especially pronounced \citep[see Figure~2 of][]{Garcia2014}. As we showed in Section\,\ref{previous}, a reanalysis of the old data with the new \texttt{relxill} model now also finds a low inclination. A second complication of the inclination measurement already discussed in \citet{Lohfink2012a}, is that for intermediate inclinations ($\sim$45\,deg) the cut-off of the blue wing of the iron line, which provides the best constraint on the inclination of the accretion disk, coincides with the Fe K$\beta$ line, possibly leading to a false association of this energy as the end of the blue wing of the line. The exact modeling of the highly ionized emission lines (Fe XXV, FeXXVI) adds additional complexity to this region of the spectrum. Future high-resolution observations with, for example, the microcalorimeter on board \textit{Astro-H}, have the potential to answer this question once and for all.

\subsection{The nature of the flux variability}
In our initial exploration of the X-ray spectrum (section \ref{res_ccd}) we discovered that the data/model residuals of an absorbed power law, with the absorption fixed at the Galactic value, are very similar for all three \textit{XMM} pointings (Fig.~\ref{resid_powerlaw}). A similar observation has been made by \citet{Keek2015} in Mrk\,335 where the soft excess component at 0.6\,keV was found to scale with the flux, yielding a constant flux ratio. However, the persistence of the fractional flux contributions is much more apparent in Fairall~9. This is most likely driven by the high black hole mass of Fairall~9 and its aborption-free spectrum, simplifying a detection significantly. The observed behavior has several implications for our understanding the X-ray spectra and spectral variability. Our analysis demonstrates that the relative contribution of each individual model component at a given energy remains constant. The number and nature of the components also remains the same. This implies that, the geometry of the X-ray emitting region cannot have changed within the time spanned by the observations. Indirectly, the accretion rate must remain approximately constant as drastic changes in the accretion rate would cause changes in the geometry. 

To gain further insight into the geometry, we model the simultaneous observations XMME and NU above 2\,keV (to avoid the complications of the soft excess) with a lamp-post geometry\footnote{\texttt{relxill} was replaced with \texttt{relxilllp} to perform this fit and $a$ and $i$ were kept fixed at their multi-epoch best fit values.}, where the corona is approximated by a point source above the disk at heigh $h$. We find that the corona would be at a height of $11_{-3}^{+7}\,\text{R}_\text{g}$ above the accretion disk. The time difference ($\sim$\,10\,days) between the two observations XMMC and XMMD combined with the fact that we do not see a significant lag of the cold reflection flux behind the continuum flux, allows us to place an upper limit on the distance to the cold reflector of 690\,$R_\text{g}$. This indicates that the origin of the cold reflection is the outer part of the accretion disk and not a more distant structure such as the torus. A radius of $690\,R_\text{g}$ lies well within the dust sublimation radius of Fairall~9 ($\sim 26000 \,R_\text{g}$) and our finding therefore agrees well the claim of \citet{Gandhi2015} that the dust sublimation radius represents an upper limit to iron emitting region. 

Having set out in this paper to find the driver of the relatively rapid variability seen in the X-ray monitorings of Fairall~9, the question now is: What is causing the variability we observe in the X-ray spectra of this AGN? From our results, it is clear that the variability of the individual components is driven by the variability of the continuum (power law). Both the cold and the ionized reflection components respond to these changes in the continuum, as does the soft excess. The question of the nature of the variability is therefore transferred to the question of what causes the continuum variability. Current models connect the continuum variability to changes of the accretion flow \citep{Gleissner2004,Uttley2004}, for example through inward propagating fluctuations of the accretion rate \citep{Lyubarskii1997,Churazov2001,Vaughan2003,Arevalo2006}. But to fully understand how these accretion flow changes are connected to changes in the coronal emission will require a better understanding of the detailed physics governing the corona.
   
\subsection{The nature of the soft excess}
Our detailed spectral analysis of the multi-epoch data revealed an additional continuum-like component in the spectra above the blurred reflection. In section \ref{soft} we found that a simple, steep power law describes this component well. This component varies together with the bulk of the continuum emission. Its strength with respect to the primary continuum does not change, similar to what is seen for the reflection components. The spectral shape and variability behavior (it must vary on relatively fast time scales, which are shorter than the time difference between the observations) suggest that Comptonization could give rise to the observed power law. One possibility is that the primary corona is non-uniform in its temperature or optical depth and this gives rise to two seemingly distinct but coordinated varying components. One plausible explanation for this would be in-homogeneous heating of the corona as, for example, caused by different magnetic field strengths at different radii. Another possibility is that this Comptonization occurs in an area that is spatially separate from the primary Comptonization, e.g. on the surface of the innermost part of the accretion disk.

The need for an additional Comptonization component to model the broad-band \textit{XMM} and \textit{NuSTAR} spectra has also been found in Ark\,120 \citep{Matt2014}, and NGC\,5548 \citep{Mehdipour2015}. Both analyses assume the two Comptonization components are coupled, with the first corona upscattering the accretion disk photons and the second corona Comptonizing the already Comptonized photons leaving the first corona. Such a scenario was outlined in \citet{Done2012}. Two spatially separate but connected Comptonization components, however would be problematic for the case of Fairall~9. To match the observed constancy of the flux ratios, for example, a doubling in the flux output of the first corona would need to lead to a doubling of the flux output of the second corona as well. A doubling in the seed photon flux however does usually not lead to a doubling of the coronal emission; in fact it would lead to a significant cooling of the second corona. The only way to ensure the flux ratios remain constant, would be a fine-tuned heating of the second/primary corona. Therefore for Fairall~9 a non-uniform corona is more likely. It is important to note here that a separate Comptonization component in the soft X-ray would naturally also extend into the UV/optical bands. This provides another means of verifying the presence of an additional component, as a single temperature Comptonization component extending from the optical to the soft X-rays would predict a significant correlation of the bands at zero lag. This is called into question by the discovery of notable lags between the X-ray emission and the UV/optical bands \citep{Alston2012,Cameron2012,McHardy2014,Edelson2015}. Notably these lags are seen between all X-ray energies and the UV/optical emission, not just the soft X-rays, with no strong correlation seen at zero lag. Usually these lags are associated with the thermal reprocessing of X-ray coronal emission on the optical/UV emitting regions of the accretion disk. Possibly, multi-zone Comptonization, for example on the outer layers of the accretion disk (with multiple temperatures) could mimic a similar effect. 

For Fairall~9 previous work also suggests the presence of thermal reprocessing in UV/optical \citep{Lohfink2014}. We will discuss the timing and UV-X-ray insights from the new data in a forthcoming paper. 

\acknowledgments
\section*{Acknowledgments}
We thank the referee for their comments that have helped to improve the clarity of the paper. AL and ACF acknowledge support from ERC Advanced Grant FEEDBACK. CSR thanks the Simons Foundation Fellows Program (US) and the Sackler Fellowship Program (Cambridge) for support. This work made use of data from the \textit{NuSTAR} mission, a project led by the California Institute of Technology, managed by the Jet Propulsion Laboratory, and funded by the National Aeronautics and Space Administration. This research has made use of the \textit{NuSTAR} Data Analysis Software (NuSTARDAS) jointly developed by the ASI Science Data Center (ASDC, Italy) and the California Institute of Technology (USA). This research has made use of data obtained from the \textit{Suzaku} satellite, a collaborative mission between the space agencies of Japan (JAXA) and the USA (NASA).

\bibliographystyle{apj}

\end{document}